\documentclass[a4paper,10pt]{article}

\usepackage{amsmath,amssymb}
\usepackage{amsthm} 
\usepackage[a4paper,top=2cm,bottom=2cm,left=2cm,right=2cm,marginparwidth=2cm]{geometry}

\usepackage{hyperref}
\usepackage{amsfonts}
\usepackage{verbatim}
\usepackage{enumitem}
\usepackage{stmaryrd}
\usepackage{graphicx}
\usepackage{siunitx}
\usepackage{svg}
\usepackage{xr}
\usepackage{booktabs}
\usepackage{bm}

\usepackage[style=numeric, sorting=none, abbreviate=true, giveninits=true, terseinits=true,maxnames=10, doi=false, isbn=false, url=false, eprint=true, date=year]{biblatex}
\DeclareFieldFormat{pages}{#1}
\renewbibmacro{in:}{%
  \ifentrytype{article}
    {}
    {\bibstring{in}%
     \printunit{\intitlepunct}}}

\usepackage{authblk}

\author[1]{Valeria Giunta} 
\author[2]{Thomas Hillen}
\author[3]{Mark A Lewis}
\author[4]{Jonathan R Potts}

\affil[1]{Department of Mathematics, Swansea University, Computational Foundry,
Bay Campus, Swansea SA1 8EN, UK}
\affil[2]{Department of Mathematical and Statistical Sciences, University of Alberta,
Edmonton, AB T6G 2G1, Canada}
\affil[3]{Department of Mathematics and Statistics and Department of Biology,
University of Victoria, PO Box 1700 Station CSC, Victoria, BC, Canada}
\affil[4]{School of Mathematical and Physical Sciences, Hicks Building, Hounsfield
Road, University of Sheffield, UK, S3 7RH}

\date{}

\title{A phylogeny of biological patterns formed by nonlocal advection}

\begin{document}

\maketitle
\begin{abstract}
From tumour invasion to cell sorting and animal territoriality, many biological systems rely on nonlocal interactions that drive complex spatial organisation. Partial differential equations (PDEs) with nonlocal advection are increasingly recognised as powerful tools for capturing such phenomena. However, most research has focused on one-dimensional domains, leaving their two-dimensional behaviour largely unexplored. Here, we present a  detailed numerical study of the patterns formed by these systems on 2D domains. Depending on the underlying mechanisms, a wide variety of spatial patterns can emerge -- including segregated clusters, stripes, volcanos, and polygonal mosaics -- many of which have been observed in natural systems. By systematically varying model parameters, we classify the links between emergent patterns and their underlying movement mechanisms. In comparing these patterns with empirical observations, we show how this modelling framework can help reveal possible mechanisms of self-organisation in various situations within the life sciences, from ecology and developmental biology to cancer research.
\end{abstract}
\section{Introduction}
The phenomenon of pattern formation has long fascinated scientists from many disciplines \cite{kauffman1992origins}. From the coordinated movements of flocks of birds and schools of fish to the formation of honeycomb lattices in beehives, nature is full of striking examples of order emerging from disorder. 
More remarkably, similar patterns occur in very different systems--for example, the sorting of cells into segregated patterns during tissue growth \cite{katsunuma2016synergistic} and the segregation of animal populations into spatially-distinct territories \cite{adams2001approaches}--raising the question of whether this similarity suggests common features in underlying mechanism. 
Indeed, despite the diversity of the agents involved, these structures all emerge from the same fundamental principle: at the microscopic level, individuals (particles, cells or animals) interact according to simple local rules (for example, attraction or repulsion); then, when these interactions occur over many individuals, they give rise to  macroscopic emergent patterns. Consequently, self-organisation has become a pillar of mathematical biology, underpinning many models used to understand pattern formation and collective dynamics \cite{murray2003spatial}.

Recently, nonlocal advection-diffusion models have emerged as a powerful framework for capturing the spatial organisation of interacting populations \cite{topaz2006nonlocal,painter2010impact,potts2016memory,carrillo2019population, hillen2020nonlocal}. These formulations take into account the effect of spatially nonlocal sensing of other organisms on organism movement , which could be mediated through sight, smell, hearing, or touch. Biological examples of nonlocal interactions are abundant. In ecology, they range from the relatively proximate sensing of swarming insects and flocking birds \cite{sumpter2010collective} to longer-range interactions between vertebrate species \cite{latombe2014spatio, vanak2013moving}. It is even possible to place indirect {\em local} interactions into the nonlocal advection-diffusion framework, for example those mediated by memory or marks on the landscape \cite{potts2016memory, potts2016territorial}. In cell biology,  by extending filopodia (essentially tentacle-like protrusions), cells can sense the presence of others at distances of several cell diameters \cite{kornberg2014cytonemes,chen2020mathematical}. Such non-local sensing is believed to be a key process behind various observed phenomena such as cell sorting, cellular aggregations, and chase-and-run dynamics \cite{painter2024biological,pal2025nonlocal}. 

Mathematically, nonlocal sensing of all these kinds can be captured by a nonlocal advection term in a partial differential equation. These nonlocal advection terms can drive spontaneous pattern formation in both single species
\cite{carrillo2019aggregation} and multi-species \cite{potts2019spatial} systems. Existing studies have documented a wide range of patterns that emerge from such systems, including stationary aggregations, segregation patterns, time-oscillating aggregates, travelling waves, period-doubling routes to chaos, and chase-and-run dynamics \cite{potts2019spatial,giunta2022local,jewell2023patterning,painter2024variations}. These have been uncovered using analytic techniques such as linear stability \cite{potts2019spatial,jewell2023patterning}, weakly non-linear analysis \cite{giunta2024weakly}, Crandall-Rabinowitz bifurcation theory \cite{hillen2020nonlocal, liu2023biological}, and energy approaches \cite{carrillo2021phase,giunta2022detecting, bailo2024aggregation}, as well as through numerical simulations \cite{bailo2018fully,painter2024variations}. 

In one spatial dimension, these approaches have revealed quite rich pattern formation properties from systems with only two species \cite{potts2019spatial,giunta2022detecting,giunta2024weakly,painter2024variations}, involving various configurations of aggregation amongst and between species, partial and full segregation between species, as well as perpetual oscillatory and chase-and-run dynamics. However, most of these studies have focused on one spatial dimension, with the effect of higher dimensions being less explored, despite having qualitatively different patterning properties \cite{jewell2023patterning}. 

Here, we analyse numerically the patterns that form from a multi-species system of advection-diffusion equations in two spatial dimensions, where the nonlocal advection terms may model either attraction or repulsion, whether between individuals of the same population (self-interaction) or between those of different populations (mutual interaction).  We investigate the range of spatial patterns that can form within this system, focusing particularly on a two-species system but also investigating certain patterns for more than two species. Our aim is to make steps toward a categorisation of these emergent patterns and how the underlying processes drive changes in emergent patterns. The resulting categorisation takes a form analogous to a phylogenetic tree from evolutionary biology, but in our case the trunk is the spatially-homogeneous steady state (i.e. no patterns) and the branching points represent bifurcations from one pattern to another, driven by changes in one or more of the parameter values, rather than by changes in genes.

Having constructed this `phylogeny', we examine which of the observed patterns correspond to those found in natural systems. The aim here is to demonstrate how our model might help elucidate the collective motion mechanisms behind the formation of observed patterns. In situations where we were unable to identify biological patterns similar to those discovered numerically, this gives rise to the question of why not: has biology evolved to avoid these mechanisms or have we simply not observed them, despite such patterns perhaps being present in the empirical world? Overall, our study helps shed light on the variety of patterns that can form purely as a result of the type of non-local advection processes that are ubiquitous across the biological sciences.
\section{The Model}
Let $u_1(x,t),\dots,u_N(x,t)$ each denote the density of a population of organisms at time $t$. The purpose of our study is to investigate numerically the following system 
\begin{equation}\label{eq:model}
	\frac{\partial u_{i}}{\partial t} = D_i \Delta u_i +\nabla\cdot\left( u_i \sum_{j=1}^N \gamma_{ij} \nabla (K_{ij} \ast u_j) \right), \qquad i=1,\dots,N, 
\end{equation}
defined on a $2$D spatial domain. 
The constant $D_i>0$ is the diffusion coefficient of species $i$, for $i=1,\dots,N$. The functions $K_{ij}$ are zero-mean probability density functions describing the nonlocal detection of species $j$ by species $i$. We assume that the support of $K_{ij}$ is ${\cal {B}}^2_{R_{ij}}$, where ${\cal {B}}^2_{R_{ij}}$ denotes the $2$-dimensional sphere with radius $R_{ij}$. From a biological point of view, the constant $R_{ij}$ denotes the detection interval within which the species $i$ is able, or willing, to interact with species $j$. To ensure that the solution remains positive and exists for all time, we assume that $K_{ij} $ is a twice-differentiable, with $\nabla K_{ij} \in L^\infty$ \cite{giunta2023positivity}. The function $K_{ij} \ast u_j$ denotes a convolution operator defined as 
\[ K_{ij}\ast u_j (x) = \int_\Omega K_{ij}(x-y) u_j(y) dy.\]
For simplicity with nonlocal boundary conditions, we consider solutions on a torus. We let $\Omega=\mathbb{T}$, the torus defined by $[0,L_1]\times[0,L_2]$ with periodic boundaries, for constants $L_1, L_2>0$. These periodic boundary conditions imply the conservation of the total mass of each population over time. Indeed, if we denote by $m_i$ the total mass of population $i$, we have
\begin{equation}\label{eq:mi}
     \frac{d m_i}{dt} = \frac{d}{dt} \int_\mathbb{T} u_i d\mathbf{x} = 0, \, \text{ for all } i=1,2, \dots, N.
\end{equation}
As a consequence, the system has a unique steady state, $\mathbf{\bar{u}}=(\bar{u}_1, \dots \bar{u}_N)$, where $\bar{u}_i=m_i/|\mathbb{T}|$, for $i=1,\dots,N$. Here, $|\mathbb{T}|=L_1\times L_2$ is the surface area of  $\mathbb{T}$.

To date, we have gained a solid understanding of the well-posedness of System \eqref{eq:model} in every spatial dimension \cite{carrillo2020long,giunta2022local,giunta2023positivity}, including scenarios involving less regular kernels \cite{jungel2022nonlocal,carrillo2024well}. The process of pattern formation has been extensively studied on 1D spatial domains \cite{giunta2022detecting}, and we now have a set of tools that allow us to extend this analysis to higher dimensions \cite{jewell2023patterning}, overcoming the difficulties arising from the integral terms. Additionally, weakly nonlinear analysis on 1D domains has provided valuable insights into the bifurcation structure of the system beyond the linear regime, revealing a wealth of possible patterns and their stability \cite{giunta2024weakly}. Through the analysis of the energy functional, in \cite{giunta2022detecting} we have uncovered the existence of strongly modulated solutions over large regions of the parameter space. These results highlight the richness of the solution landscape of the system. Not only do these strongly modulated patterns often coexist with the homogeneous solution, but in certain parameter regimes they coexist with more regularly patterned solutions \cite{giunta2024weakly}, adding an additional layer of complexity to the system.

These findings demonstrate that the model in \eqref{eq:model} is capable of supporting a wide range of emergent patterns, even in the relatively simple case of 1D spatial domains and $N=2$, with an even greater range of patterns for $N>2$ \cite{potts2019spatial}. This naturally leads to the expectation that in higher spatial dimensions, such as 2D, the system will exhibit a yet broader and range of patterns. This insight would be not only interesting {\it per se}, as it would deepen our understanding of the dynamics supported by the model, but also has practical implications. By moving to two-dimensional spatial domains, we gain a more realistic representation of natural processes, opening the door to a variety of biological applications.

Given the considerable complexity of the parameter space, it is impractical to undertake a comprehensive exploration.  Instead, we take a strategic approach by selecting specific cases in which certain parameters are held constant while others are systematically varied. This method allows us to explore the role of individual parameters in shaping the dynamics of the system, and facilitates a clearer understanding of how these parameters influence pattern formation and stability.

\begin{table}[h!]
\centering
\begin{tabular}{r l @{\hspace{0.6cm}} r l @{\hspace{0.6cm}} r l}
\toprule
\textbf{No.} & \textbf{Pattern Name} & \textbf{No.} & \textbf{Pattern Name} & \textbf{No.} & \textbf{Pattern Name} \\
\midrule
1 & \textit{Homogeneous distribution} 
  & 8 & \textit{Co-Aggregate} 
  & 15 & \textit{Single Species Aggregate ($u_2$)} \\
2 & \textit{Segregated Clusters} 
  & 9 & \textit{Refuge aggregation ($u_2$)} 
  & 16 & \textit{Refuge Volcano ($u_1$)} \\
3 & \textit{Diamond Patterns} 
  & 10 & \textit{Refuge Aggregation ($u_1$)} 
  & 17 & \textit{Fenced Refuge Volcano ($u_1$)} \\
4 & \textit{Stripes} 
  & 11 & \textit{Splash Pattern ($u_2$)} 
  & 18 & \textit{Chase-and-run} \\
5 & \textit{Ridged Stripes} 
  & 12 & \textit{Splash Pattern ($u_1$)} 
  & 19 & \textit{Oscillations} \\
6 & \textit{Mixed Stripes} 
  & 13 & \textit{Co-Aggregate with halo} 
  & & \\
7 & \textit{Diagonal Stripes} 
  & 14 & \textit{Cluster in Volcano} 
  & & \\
  & & & \textit{with halo ($u_1 - u_2$)} & & \\
\bottomrule
\end{tabular}
\caption{Descriptive names assigned to the observed spatial patterns. The numbers correspond to the panels in Figure~\ref{fig:legend}.}
\label{tab:pattern_names}
\end{table}
By varying the parameters in a controlled manner, we aim to identify critical thresholds and transitions that can lead to qualitatively different behaviours in the model. This targeted investigation not only improves our understanding of the system, but also helps to uncover the relationships between parameters that may not be apparent in a broader analysis.

\section{Analysing the role of the interaction parameters $\gamma_{ij}$}\label{sec:gammaij}
The model has a large number of parameters, 
each of which might influence the emergent patterns. However, we will mainly focus on the study of the nonlocal interaction parameters, $\gamma_{ij}$, and aim to uncover their influence on pattern formation. To ensure that our analysis on the interaction parameters is meaningful, we fix all the other parameters of the system to equal values, specifically setting $D_i =D_j$, $m_i=m_j$, and $K_{ij} = K_{ij} $, for all $i,j$. This creates a symmetric scenario that allows us to isolate the effect of the interaction parameters on the emergent patterns. We also predominantly analyse the system with $N=2$ populations (with a small exploration of $N>2$ in certain cases). This choice simplifies the analysis while still capturing the main features of the system. 
Studying more populations could introduce additional complexity and potentially obscure the primary role played by the interacting parameters in shaping the emergent patterns. 
By focusing on two populations, we maintain the clarity of our results and also provide a foundation for future investigations of multi-species dynamics.

\begin{figure}[h!] 
\centering
         {\includegraphics[width=1\textwidth]{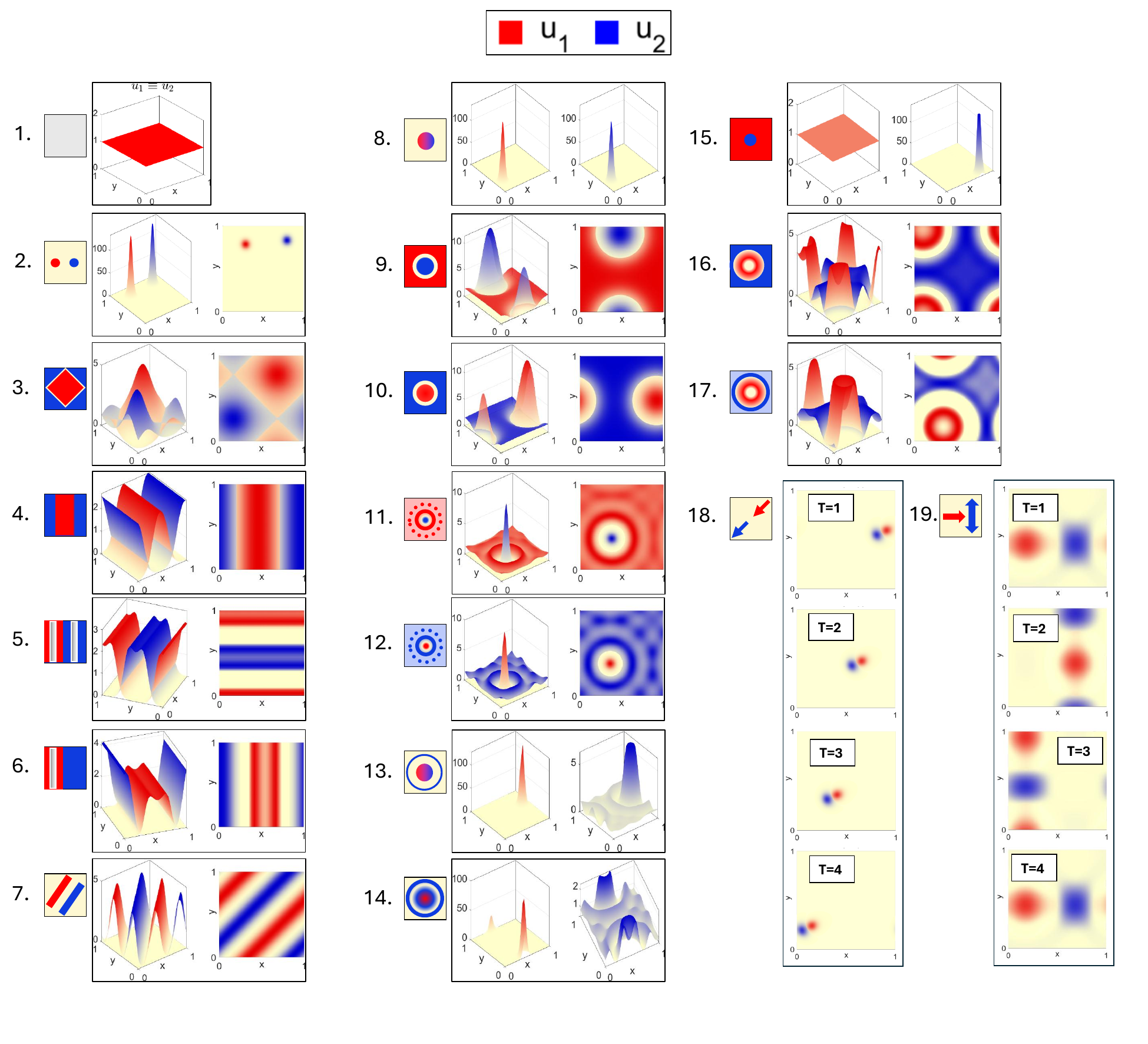}} 
         \caption{List of stationary and non-stationary numerical solutions obtained by solving System \eqref{eq:model}, with $K_{ij}$ defined as in \eqref{eq:kernel}, for different values of the nonlocal interaction parameter $\gamma_{ij}$, for $i,j=1,2$. The other parameter values are: $N=2$, $D_1=D_2=1$, $R_{11}=R_{12}=R_{21}=R_{22}=0.3$, $m_1=m_2=1$, $L_1=L_2=1$. Each of these patterns is associated with an icon shown on the left of the corresponding numerical solution. These icons will be used in subsequent figures to reference the type of solution.}
         \label{fig:legend}
\end{figure}
We fix $N=2$, $D_1=D_2=1$, $m_1=m_2=1$, and work on the spatial domain $[0,L_1] \times [0, L_2]$, with $L_1=L_2=1$. For this choice of parameter values, the homogeneous steady state is $\mathbf{\bar{u}}=(\bar{u}_1, \bar{u}_2)= (1,1)$. We use the following kernel
\begin{equation}\label{eq:kernel}
    K_{ij}(x,y)=\begin{cases}
         \frac{\pi}{R_{ij}^2(\pi^2-4)}\left( 1+\cos\left( \frac{\pi}{R_{ij}}\sqrt{x^2+y^2} \right) \right), & \text{ if } x^2+y^2 \leq R_{ij}^2, \\
         0, & \text{ otherwise,}
    \end{cases}
   \end{equation}
and fix $R_{11}=R_{12}=R_{21}=R_{22}=0.3$.

In our numerical investigation, we selected a number of cases, which are listed and commented on below. We started with the simplest scenario, where $\gamma_{11} = \gamma_{22}$ and $\gamma_{12} = \gamma_{21} $, and we then included more complex cases in which the interaction parameters $\gamma_{ii}$ and $\gamma_{ij}$ are decoupled, to understand their specific role in the pattern formation process.

Our numerical investigation has shown that by varying the interaction parameters, $\gamma_{ij}$, the system gives rise to various classes of emergent patterns, listed in Figure \ref{fig:legend}, which can be either stationary (Solutions from 1 to 17 in Figure \ref{fig:legend}) or non-stationary (Solutions 18 and 19 in Figure \ref{fig:legend}). We assign an icon to each of these patterns, shown to the left of the corresponding numerical solution in Figure \ref{fig:legend}, as a way of referencing the solution type in subsequent figures. To further aid our discussion, we assign a descriptive name to each of the observed solutions.  These names are collected in Table \ref{tab:pattern_names}, which links each number (as shown in Figure \ref{fig:legend}) to its corresponding pattern name. 

\subsection{Case 1: Symmetric interactions (${\gamma_{11} = \gamma_{22}} $ and ${\gamma_{12} = \gamma_{21} }$) }

\noindent
In this first scenario, we set $\gamma_{11} = \gamma_{22}$, denoted by $\gamma_{ii}$, and set $\gamma_{12} = \gamma_{21}$, denoted by $\gamma_{ij}$. Our goal is to understand the effect of the interaction parameters, $\gamma_{ii}$ (self-interaction) and $\gamma_{ij}$ (mutual interaction), on the stability of the homogeneous steady state $\mathbf{\bar{u}}= (1,1)$ and on the properties of the emergent solutions when $\mathbf{\bar{u}}$ is unstable. Linear stability analysis, whose details can be found in the Supplementary Information (SI), reveals that for $-1.25\lesssim\gamma_{ii}\lesssim 50$ the homogeneous steady state is stable if and only if $\gamma_{ii} - 50\lesssim\pm\gamma_{ij} \lesssim \gamma_{ii} + 1.25$, while it is unstable for $\gamma_{ii}\gtrsim50$ or $\gamma_{ii}\lesssim-1.25$. For numerical simulations, we restricted $\gamma_{ii}$ and $\gamma_{ij}$ to integer values within the range $[-5, 5]$. Within this parameter range and for integer values of the interaction parameters, we have the following stability conditions: for $\gamma_{ii} \geq -1$, the homogeneous steady state is stable if and only if $\gamma_{ij}^2 \leq (1+\gamma_{ii})^2$; for $\gamma_{ii} < -1$, $\mathbf{\bar{u}}$ is unstable for all $\gamma_{ij}$ (see Equations (S8) and (S9) in SI).
\begin{figure}[h] 
       \centering  {\includegraphics[width=0.5\textwidth]{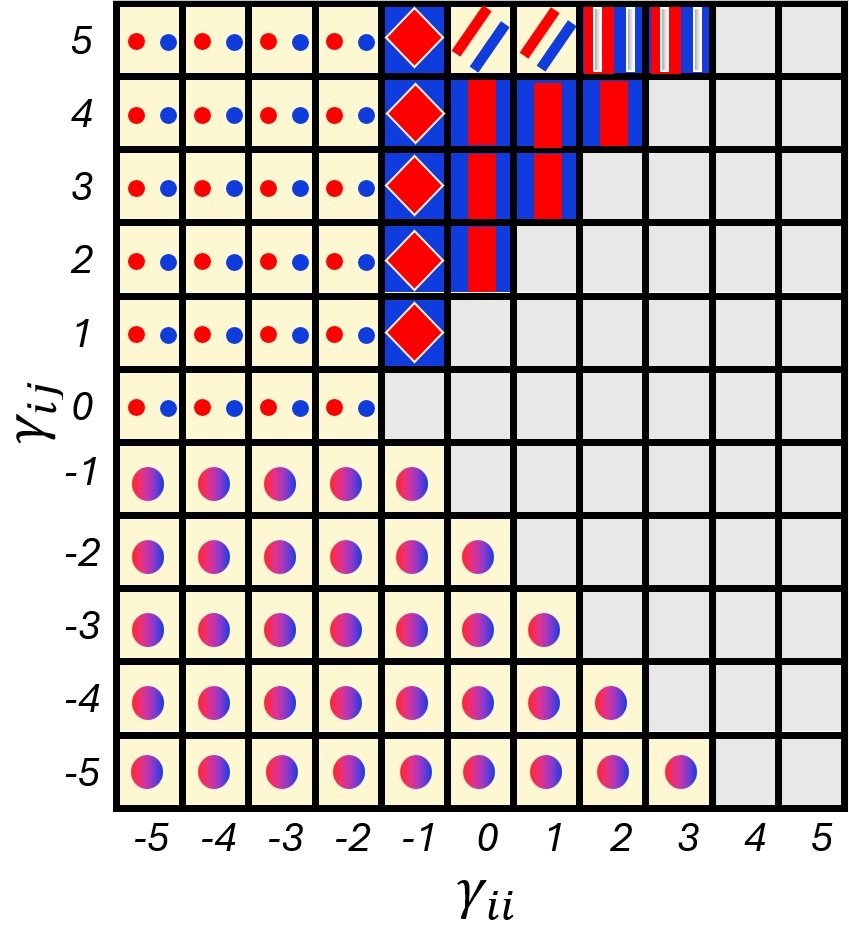}}         \caption{\textbf{\textit{Case 1: Symmetric interactions.}} Pattern selection across the interaction parameter space $(\gamma_{ii},\gamma_{ij})$, where  $\gamma_{ii}=\gamma_{11} = \gamma_{22} $ and ${\gamma_{ij} =\gamma_{12} = \gamma_{21}}.$ We classify the numerical solutions obtained by solving System \eqref{eq:model}, with $K_{ij}$ given in \eqref{eq:kernel}, for each pair of integer values $(\gamma_{ii},\gamma_{ij}) \in [-5,5] \times [-5,5]$. The other parameter values are as in Figure \ref{fig:legend}. By varying the interaction parameters, we find a number of solutions. Each solution is associated with an icon indicating its qualitative property. The legend of the icons is given in Figure \ref{fig:legend}.}
         \label{fig:Case1}
\end{figure}

In all simulations, the initial condition is a small random perturbation of the homogeneous steady state. The results of our numerical investigation are summarised in Figure \ref{fig:Case1}, where we show the state of the system after the transient dynamics have subsided. When $\gamma_{ij}<0 $ (mutual attraction) and $\gamma_{ii} <0 $ (self-attraction), the system evolves towards a stationary \textit{Co-Aggregate} (see e.g. Figure \ref{fig:Case1}, $(\gamma_{ii},\gamma_{ij})=(-5,-5)$). This solution persists in the self-avoidance case, $\gamma_{ii}> 0$, as long as $\gamma_{ij}$ is sufficiently far below zero to counteract the self-avoidance (see $(\gamma_{ii},\gamma_{ij})=(3,-5)$). However, if $\gamma_{ii}$ becomes sufficiently high, so that $\gamma_{ij}^2 \leq (1+\gamma_{ii})^2$, then we drop in the stability region and patterns fail to form (see $(\gamma_{ii},\gamma_{ij})=(4,-5)$).

On the other hand, if $\gamma_{ij}>0 $ (mutual repulsion), segregation patterns form in the instability region $\gamma_{ij}^2>(1+\gamma_{ii})^2$. When there is self-attraction, i.e. $\gamma_{ii}<0$, the populations arrange into \textit{Segregated Clusters}. These stripes can either be parallel (\textit{Stripes}) to one of the axes, as in the case of $(\gamma_{ii},\gamma_{ij})=(1,3)$, or transversal (\textit{Diagonal Stripes}) to one of the axes, as in the case of $(\gamma_{ii},\gamma_{ij})=(1,5)$.
For sufficiently high values of self- and mutual-repulsion, there is a further transition from \textit{Stripes} to \textit{Ridged Stripes} (e.g. compare $(\gamma_{ii},\gamma_{ij})=(2,4)$ with $(\gamma_{ii},\gamma_{ij})=(3,5)$). As the mutual repulsion increases, the stripes become taller and ridges appear at high values of self-repulsion. These ridges therefore seem to be the result of strong repulsive forces preventing the formation of high densities.

\subsection{Case 2: No self-interaction in $\mathbf{u_1}$  (${\gamma_{11} = 0 }$)}

\noindent
We now fix $\gamma_{11} = 0 $ (so that $u_1$ has no self-interaction) and vary $\gamma_{22}$, $\gamma_{12}$, and $\gamma_{21}$. Our analysis focuses on three values of $\gamma_{22}$, namely $-1$, $0$, and $1$.  Linear stability analysis (see SI) shows that patterns form only when $\gamma_{12} $ and $\gamma_{21} $ have the same sign (see Equations (S10)-(S12)). Also, increasing values of $\gamma_{22}$ favour the stability of the homogeneous steady state. That is the stability region expands as $\gamma_{22}$ increases from $-1$ to $1$ (compare (S10)-(S12)). From a biological perspective, this is expected, as stronger self-repulsion reduces the likelihood of aggregation by driving the dispersion of individuals in space.
\begin{figure}[h!] 
       \centering  {\includegraphics[width=1\textwidth]{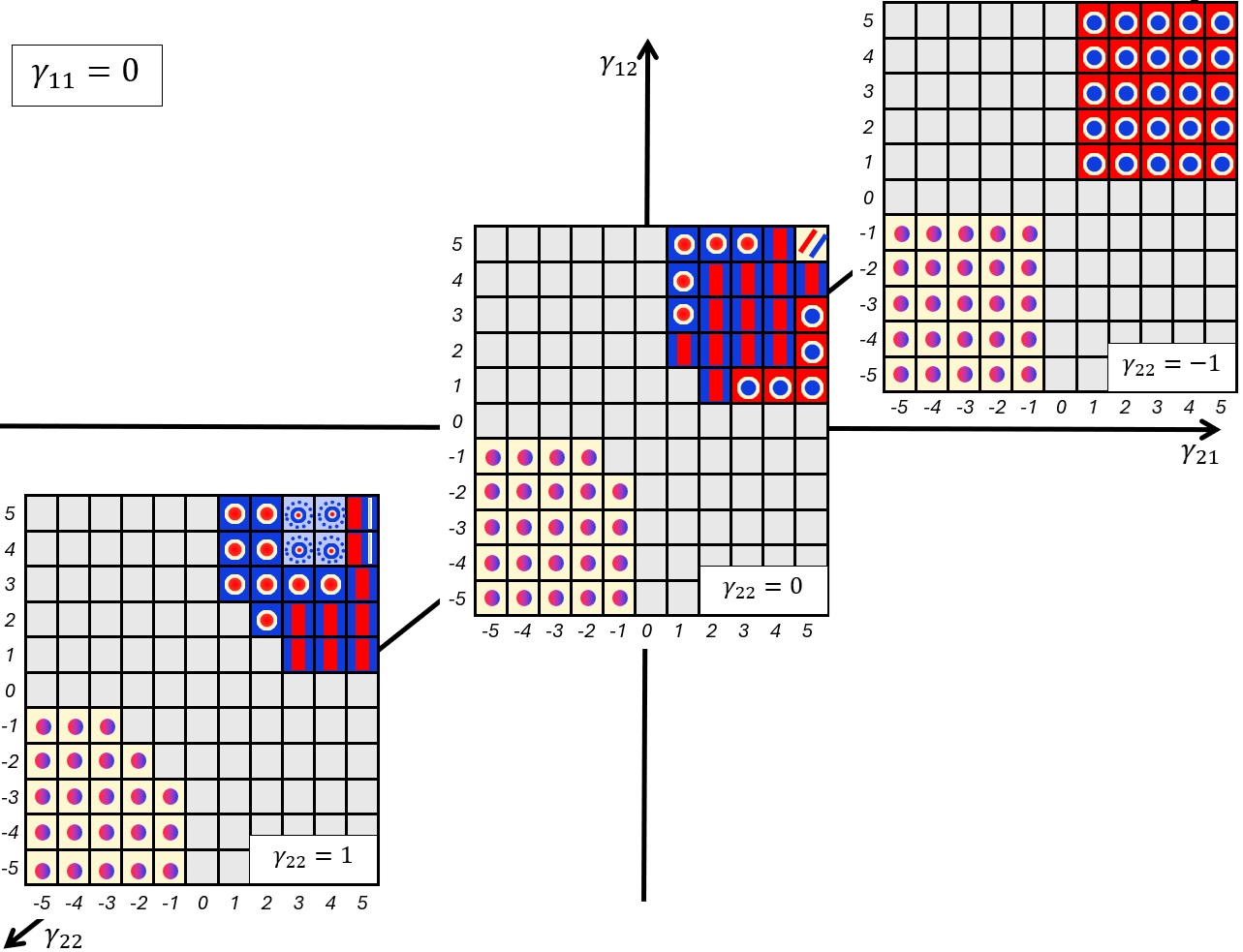}} 
         \caption{\textbf{\textit{Case 2: No self-interaction in $\mathbf{u_1}$}}. Pattern selection for $\gamma_{11}=0$ across the interaction parameter space $(\gamma_{21},\gamma_{22},\gamma_{12})$. We show the class of the numerical solutions obtained by solving System \eqref{eq:model}, with $K_{ij}$ given in \eqref{eq:kernel}, for three values of $\gamma_{22}$ ($\gamma_{22}=-1$ (right), $\gamma_{22}=0$ (center), $\gamma_{22}=1$ (left)), and for each pair of integer values $(\gamma_{21},\gamma_{12}) \in [-5,5] \times [-5,5]$. The other parameter values are as in Figure \ref{fig:legend}. By varying the interaction parameters, we find a number of solutions. Each solution is associated with an icon indicating its qualitative property. The legend of the icons is given in Figure \ref{fig:legend}.}
         \label{fig:Case2}
\end{figure}

We numerically solved System \eqref{eq:model}, with $K_{ij}$ given in Equation \eqref{eq:kernel}, for each pair of integer values $(\gamma_{21},\gamma_{12}) \in [-5,5] \times [-5,5]$ and the results are shown in Figure \ref{fig:Case2}. We will start the discussion of our results from the case $\gamma_{22} = 0$ (Figure \ref{fig:Case2}, middle grid) and focus on the instability region 
$\gamma_{12}\gamma_{21}\geq 2$ (see Equation (S10)). When $\gamma_{12},\gamma_{21}<0$ (mutual attraction), the system forms a \textit{Co-Aggregate}, mirroring what we observed in Case 1 for $\gamma_{ij}<0$. When $\gamma_{12}, \gamma_{21} >0$ (mutual avoidance), segregation patterns emerge. If both $\gamma_{12}$ and $\gamma_{21}$ are similar in magnitude, these segregation patterns take the form of \textit{Stripes} (see, for example, $(\gamma_{21},\gamma_{12})=(3,3))$ or \textit{Diagonal Stripes} (see, for example, $(\gamma_{21},\gamma_{12})=(5,5)$). Based on this result and the observations from the previous case (see Figure \ref{fig:Case1}, with $(\gamma_{ii}, \gamma_{ij}) = (1, 5)$), \textit{Diagonal Stripes} seem to emerge when the mutual avoidance parameters have similar magnitude and reaches sufficiently high levels. But if they differ sufficiently, we observe the formation of \textit{Refuge Aggregations} whereby one of the populations aggregates whilst the other spreads out evenly across the rest of space (see, for example, $(\gamma_{21},\gamma_{12})=(1,5)$, where $u_1$ aggregates and $u_2$ spreads out, and $(\gamma_{21},\gamma_{12})=(5,1)$, where $u_2$ aggregates and $u_1$ spreads out). For example, if the repulsion of $u_2$ away from $u_1$ is high compared to the repulsion of $u_1$ from $u_2$, $u_2$ aggregates tightly to avoid $u_1$. On the other hand, $u_1$'s repulsion towards $u_2$ forces it to spread out through the rest of space not occupied by $u_2$.

We now turn to the case $\gamma_{22} = -1 $ (Figure \ref{fig:Case2}, right grid), so that $u_2$ exhibits self-attraction, and analyse the solutions emerging in the instability region
$\gamma_{12}\gamma_{21} \geq 1$ (see Equation (S11)). If $\gamma_{12}, \gamma_{21}<0$ (mutual attraction), \textit{Co-Aggregates} emerge. On the other hand, if $ \gamma_{12}, \gamma_{21} >0$ (mutual repulsion), \textit{Refuge Aggregations ($u_2$)} emerge, with $u_2$ forming an aggregation and $u_1$ spreading out in the remaining space. Comparing this to the case with $\gamma_{22} = 0$ (Figure \ref{fig:Case2}, middle grid), the inclusion of an asymmetry in relative self-attraction between $u_1$ and  $u_2$ means that \textit{Stripes}, which are quite symmetric in appearance, no longer emerge. Rather, the asymmetry in self-attraction leads to an asymmetry in patterning.

Finally, we focus on the case $\gamma_{22} = 1$ (Figure \ref{fig:Case2}, left grid), where $u_2$ exhibits self-repulsion, and analyse the solutions emerging in the instability region $\gamma_{12}\gamma_{21} \geq 3$ (see Equation (S12)). When $\gamma_{12},\gamma_{21}<0$, \textit{Co-Aggregations} emerge. When $\gamma_{12},\gamma_{21}>0$ (mutual repulsion),  segregation patterns emerge, but with notable differences compared to the previous cases. Specifically, there is a larger region of \textit{Refuge Aggregations ($u_1$)} than in the $\gamma_{22} = 0$ case. This enlargement can be attributed to the self-repulsion of $u_2$, which is not present in $u_1$. We also observe that there can be ridges on the distribution of $u_2$, when the interaction parameters becomes sufficiently high (see, for example, the \textit{Splash Pattern} emerging when $(\gamma_{21},\gamma_{12})=(4,4)$). If we increase $\gamma_{21}$ while keeping $\gamma_{12}$ fixed, we can observe striped patterns that take the form of either \textit{Stripes} (see $(\gamma_{21},\gamma_{12})=(5,3)$) or \textit{Mixed Stripes} (see $(\gamma_{21},\gamma_{12})=(5,4)$), where the distribution of $u_2$ is ridged because the self-avoidance mechanism prevents the population density from becoming too high at any point. Meanwhile, $u_1$, which has no self-avoidance, forms only stripes.

\subsection{Case 3: Self-repulsion in $\mathbf{u_1}$ (${\gamma_{11}=1}$)} 

\noindent
We now fix $\gamma_{11} = 1$, so that the population $u_1$ exhibits self-repulsion, and vary $\gamma_{22}$, $\gamma_{12}$, and $\gamma_{21}$. Our analysis focuses on three values of $\gamma_{22}$, namely $-1$, $1$, and $5$. Stability analysis, detailed in SI, shows that, as in the previous case, pattern formation only occurs when $\gamma_{12}$ and $\gamma_{21}$ have the same sign (see Equations (S13)-(S15) in SI). Furthermore, increasing $\gamma_{22}$ expands the stability region of the homogeneous steady state, consistent with our findings in previous cases.  
\begin{figure}[h] 
       \centering  {\includegraphics[width=1\textwidth]{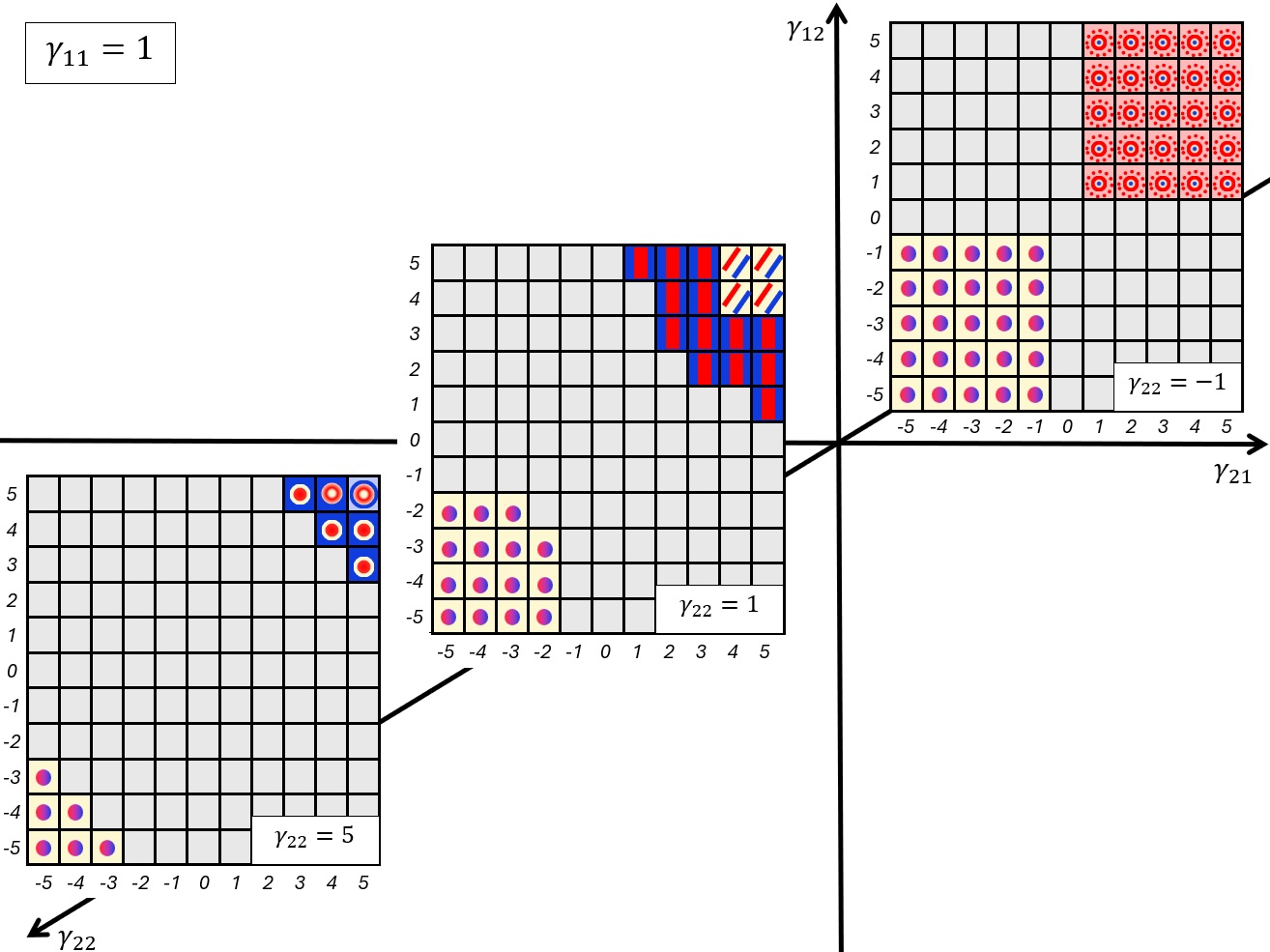}} 
         \caption{\textbf{\textit{Case 3: Self-repulsion in $\mathbf{u_1}$.}} Pattern selection for $\gamma_{11}=1$ across the interaction parameter space $(\gamma_{21},\gamma_{22},\gamma_{12})$. We show the class of the numerical solutions obtained by solving System \eqref{eq:model}, with $K_{ij}$ given in \eqref{eq:kernel}, for three values of $\gamma_{22}$ ($\gamma_{22}=-1$ (right), $\gamma_{22}=1$ (center), $\gamma_{22}=5$ (left)), and for each pair of integer values $(\gamma_{21},\gamma_{12}) \in [-5,5] \times [-5,5]$. The other parameter values are as in Figure \ref{fig:legend}. By varying the interaction parameters, we find a number of solutions. Each solution is associated with an icon indicating its qualitative property. The legend of the icons is given in Figure \ref{fig:legend}.}
         \label{fig:Case3}
\end{figure}

We conducted numerical simulations of System \eqref{eq:model} with $K_{ij}$ as defined in \eqref{eq:kernel}, for integer values of $(\gamma_{21},\gamma_{12}) \in [-5,5] \times [-5,5]$, and the results are shown in Figure \ref{fig:Case3}. Across all three values of $\gamma_{22}$, when $\gamma_{12}, \gamma_{21} < 0$ (mutual attraction), the system forms \textit{Co-Aggregates} in the instability regions.
However, the situation for $\gamma_{12}, \gamma_{21} > 0$ is more complex. We begin with $\gamma_{22} = -1$ (Figure \ref{fig:Case3}, right grid), indicating self-attraction in $u_2$. In the instability region $\gamma_{12} \gamma_{21} \geq 1$ (see Equation (S13)), \textit{Splash Patterns} emerge, where $u_2$ forms a localised aggregation due to its self-attraction, while $u_1$ disperses into the surrounding space. Notably, the distribution of $u_1$ is not homogeneous, but ridges appear. Comparing this to the case $\gamma_{11} = 0$ with $\gamma_{22} = -1$ (see Figure \ref{fig:Case2}, right grid), the solutions obtained are similar except for the ridges in $u_1$, suggesting that self-repulsion introduces spatial variations in its distribution. A similar, yet symmetric, behaviour was observed in the scenario where $\gamma_{11} = 0$ and $\gamma_{22} = 1$ (see Figure \ref{fig:Case2}, left grid, $ (\gamma_{21},\gamma_{12})=(4,4) $), with $u_1$ aggregating and $u_2$ dispersing while exhibiting ridges, another example of self-avoidance promoting ridge formation.

Next, we focus on $\gamma_{22} = 1$ (Figure \ref{fig:Case3}, middle grid), in which both populations exhibit self-repulsion of equal strength. In the instability region $\gamma_{12} \gamma_{21} \geq 5$ (see Equation (S14)), patterns emerge either in the form of \textit{Stripes} (see, for example, $(\gamma_{21}, \gamma_{12}) = (3, 3)$) or \textit{Diagonal Stripes} (see, for example, $(\gamma_{21}, \gamma_{12}) = (5,5)$). Specifically, \textit{Diagonal Stripes} emerge as the magnitude of mutual avoidance increases. This observation is consistent with those from previous cases (e.g., Figure \ref{fig:Case1}, $(\gamma_{ii}, \gamma_{ij}) = (1, 5)$ or Figure \ref{fig:Case2}, $(\gamma_{21}, \gamma_{12}) = (5, 5)$), in that strong mutual avoidance can lead to stripe or diagonal stripe solutions. 
Comparing this scenario to the case $\gamma_{11} = 0$ with $\gamma_{22} = 0$ (see Figure \ref{fig:Case2}, middle grid), we observe that self-repulsion in both populations (i.e. $\gamma_{11},\gamma_{22}>0$) prevents the emergence of \textit{Refuge Aggregations}, where one population clusters while the other disperses. 

Finally, we examine the case with $\gamma_{22} = 5$ (Figure~\ref{fig:Case3}, left grid), where $u_2$ experiences strong self-repulsion. In the instability region $\gamma_{12} \gamma_{21} > 14$ (see Equation (S15)), mutual repulsion leads to the formation of \textit{Refuge-type} solutions, with $u_1$ aggregating while $u_2$ spreads out in the surrounding space (e.g., $(\gamma_{21}, \gamma_{12}) = (4, 4)$). Compared to the case $\gamma_{22} = 1$ (Figure~\ref{fig:Case3}, middle grid), striped patterns no longer emerge, as the stronger self-repulsion in $u_2$ diminishes its capacity to aggregate. Nonetheless, mutual avoidance still supports segregation, with $u_1$ forming a distinct cluster.

As the repulsion of $u_1$ from $u_2$ increases, the \textit{Refuge Aggregate} transitions into a \textit{Refuge Volcano} configuration: for example, in moving from $(\gamma_{21}, \gamma_{12}) = (3, 5)$ to $(\gamma_{21}, \gamma_{12}) = (4, 5)$, the cluster of $u_1$ becomes more pronounced and develops a central hole. This hollowing effect is likely driven by the self-repulsion of $u_1$, which reduces density in high-concentration regions.
When mutual repulsion becomes even stronger, as in $(\gamma_{21}, \gamma_{12}) = (5, 5)$, we observe the emergence of a \textit{Fenced Refuge Volcano}. In this pattern, the strong self-avoidance of $u_2$, combined with its pronounced repulsion from $u_1$, leads to the formation of sharply defined ridges in the $u_2$ distribution that encircle the $u_1$ cluster. These ring-like structures reflect the avoidance of $u_2$ from both itself and $u_1$, effectively ``fencing in'' the volcano-shaped aggregation of $u_1$.

\subsection{Case 4: Self-attraction in $u_1$ (${\gamma_{11}=-1}$)} 

\begin{figure}[h!] 
       \centering  {\includegraphics[width=1\textwidth]{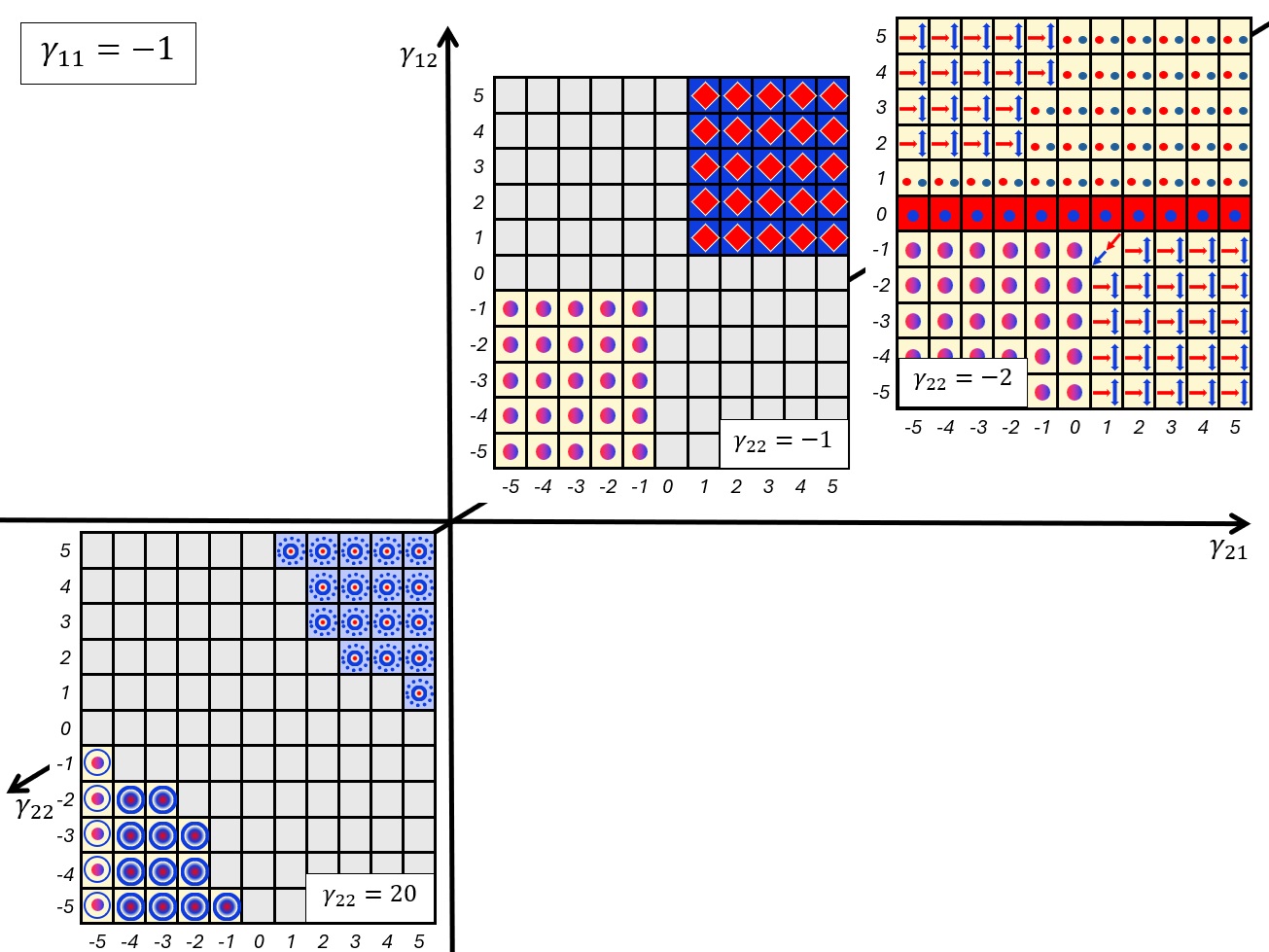}} 
         \caption{\textbf{\textit{Case 4: Self-attraction in $u_1$.}} Pattern selection for $\gamma_{11}=-1$ across the interaction parameter space $(\gamma_{21},\gamma_{22},\gamma_{12})$. We show the class of the numerical solutions obtained by solving System \eqref{eq:model}, with $K_{ij}$ given in \eqref{eq:kernel}, for three values of $\gamma_{22}$ ($\gamma_{22}=-1$ (right), $\gamma_{22}=1$ (center), $\gamma_{22}=20$ (left)), and for each pair of integer values $(\gamma_{21},\gamma_{12}) \in [-5,5] \times [-5,5]$. The other parameter values are as in Figure \ref{fig:legend}. By varying the interaction parameters, we find a number of solutions. Each solution is associated with an icon indicating its qualitative property. The legend of the icons is given in Figure \ref{fig:legend}.}
         \label{fig:Case4}
\end{figure}

\noindent
We now fix $\gamma_{11} =- 1$, so that the population $u_1$ exhibits self-attraction, and vary $\gamma_{22}$, $\gamma_{12}$, and $\gamma_{21}$. In our analysis, we focused on three values of $\gamma_{22}$, namely $-2$, $-1$, and $20$. Similar to the previous cases, we conducted both linear stability analysis and numerical simulations of System \eqref{eq:model}, with $K_{ij}$ defined as in \eqref{eq:kernel}, for integers pairs $(\gamma_{21},\gamma_{12}) \in [-5,5] \times [-5,5] $. Numerical results are shown in Figure \ref{fig:Case4}.

For $\gamma_{22} = -2$, linear stability analysis shows that the system exhibits oscillatory instability when $\gamma_{12}\gamma_{21} \leq -1$, and stationary instability for $\gamma_{12}\gamma_{21} \geq 0$ (see SI). Numerical simulations confirm this and also reveal a range of emergent patterns (see Figure \ref{fig:Case4}, right grid).
For $\gamma_{12} = 0$, i.e. $u_1$ does not interact with $u_2$, \textit{Single Species Aggregates} emerge, where $u_1$ disperses evenly over the entire spatial domain, as its self-attraction ($\gamma_{11} = -1$) is not sufficiently strong to induce clustering. Meanwhile, $u_2$ forms an aggregation, driven by its strong self-attraction ($\gamma_{22} = -2$). On the other hand, for $\gamma_{21} = 0$, meaning that $u_2$ does not interact with $u_1$, $u_2$ still forms an aggregation due to its self-attraction. However, the behaviour of $u_1$ depends on its interaction with $u_2$. If $u_1$ is attracted to $u_2$ (i.e., $\gamma_{12} < 0$), it clusters within the same region occupied by $u_2$, forming \textit{Co-Aggregates}. Conversely, if $u_1$ is repelled by $u_2$ (i.e., $\gamma_{12} > 0$), \textit{Segregated Clusters} emerge. In the region $\gamma_{12}\gamma_{21} > 0$,  we observe the usual \textit{Co-Aggregate} when $\gamma_{12}$, $\gamma_{21} <0$ (mutual attraction), and we observe \textit{Segregation Clusters} when $\gamma_{12}$, $\gamma_{21} >0$ (mutual repulsion).

In the region predicted to exhibit oscillatory behaviour ($\gamma_{12}\gamma_{21} \leq -1$), numerical simulations confirm the presence of oscillations (see, for example, Figure \ref{fig:Case4}, right grid, $(\gamma_{21},\gamma_{12}) = (2,-2)$ or $(\gamma_{21},\gamma_{12}) = (-2,2)$). Interestingly, we also identified a case of \textit{Chase-and-run} dynamics, for $(\gamma_{21},\gamma_{12}) = (1,-1)$, where $u_2$ aggregates and moves away from $u_1$, while $u_1$ forms an aggregation that actively pursues $u_2$, resulting in a persistent chase. This behaviour develops after a transient phase: initially the system exhibits oscillatory behaviour, which gradually evolves into a sustained chase-and-run motion. However,  chase-and-run dynamics do not emerge when the interaction strengths $(\gamma_{21},\gamma_{12})$ are varied (see $(\gamma_{21},\gamma_{12}) = (2,-1)$ or $(\gamma_{21},\gamma_{12}) = (1,-2)$). This phenomenon has been analysed previously in \cite{painter2024variations}, for a similar but not identical model. There, the authors showed that for the chase-and-run dynamics to remain robust under parameter variations, the sensing range of the chaser - $u_1$ in this case - must be larger than that of the runner, $u_2$. In our simulations, the sensing ranges of $u_1$ and $u_2$ are equal, thus the chase-and-run dynamics is not preserved as the mutual interaction parameters are varied.

While in some cases the initial oscillations evolve into sustained \textit{Chase-and-run} dynamics, in others they are entirely suppressed as the system stabilises into stationary segregation patterns (see for example Figure \ref{fig:Case4}, right grid, $(\gamma_{21}, \gamma_{12}) = (-5,1)$). The numerical simulations initially show oscillatory behaviour, similar to the transient phase observed in the chase-and-run case. Over time, however, these oscillations decay and the system stabilises into a stationary configuration where $u_1$ and $u_2$ segregate into distinct clusters. The emergence of these stationary segregation patterns appears to be related to the strength of $u_1$ avoidance towards $u_2$. Specifically, they are observed when this avoidance is relatively weak (for example $\gamma_{12} = 1$), suggesting that stronger repulsion may be required to sustain persistent and periodic \textit{Oscillations}.

Next, we focus on $\gamma_{22} = -1$ (Figure \ref{fig:Case4}, middle grid), in which both populations exhibit self-attraction of equal strength. Linear stability analysis predicts a stationary instability when $\gamma_{12}\gamma_{21} \geq 1$ (see Equation (S16)). Numerical simulations show that in the mutual attraction regime ($\gamma_{12}, \gamma_{21} < 0$), \textit{Co-Aggregates} emerge. In the mutual repulsion regime ($\gamma_{12}, \gamma_{21} > 0$), \textit{Diamond Patterns} emerge, representing a shift from the symmetric striped patterns seen in earlier cases (compare with Figure \ref{fig:Case2}, middle grid, and Figure \ref{fig:Case3}, middle grid). These rhombic clusters arise from the combined effect of self-attraction in $u_1$ and $u_2$ and their mutual repulsion, producing structures that lie between round clusters and fully developed striped patterns. 

Finally, we analyse the case where $\gamma_{22} = 20$, representing a scenario where $u_2$ exhibits strong self-repulsion. Linear stability analysis predicts that the homogeneous steady state becomes unstable when $\gamma_{12} \gamma_{21} \geq 5$ (see Equation (S17)). Numerical simulations confirm this prediction (see Figure \ref{fig:Case4}, left grid) and show that in the instability region, mutual attraction ($\gamma_{12}, \gamma_{21} < 0$) leads to the formation of aggregation patterns. In this region, $u_2$ clusters close to $u_1$, but also forms surrounding ridges or waves, resembling a \textit{Halo}, due to its strong self-repulsion (see e.g. Figure \ref{fig:Case4}, left grid, $(\gamma_{21}, \gamma_{12}) = (-5,-1)$). The combination of the strong self-repulsion in $u_2$ and its attraction to $u_1$ can lead to the formation of a \textit{Cluster in Volcano with Halo}, where $u_1$ and $u_2$ form a cluster, but $u_2$ also develops a distinct \textit{hole} at its centre (see e.g. $(\gamma_{21}, \gamma_{12}) = (-1,-5)$). As the attraction of $u_2$ towards $u_1$ increases, the attraction gradually overcomes $u_2$'s self-repulsion, causing the hole to disappear and giving rise to a \textit{Co-Aggregate with Halo} (compare the cases $(\gamma_{21}, \gamma_{12}) = (-1,-5)$ and $(\gamma_{21}, \gamma_{12}) = (-5,-5)$).
In the mutual repulsion regime ($\gamma_{12}, \gamma_{21} > 0$), \textit{Splash Patterns} emerge, with $u_1$ forming a distinct cluster, while $u_2$ disperses into the remaining space, developing ridges as a result of its strong self-avoidance.

\subsection{Summary of Pattern Formation and Parameter Influence}

\begin{figure}[h!] 
       \centering  {\includegraphics[width=1\textwidth]{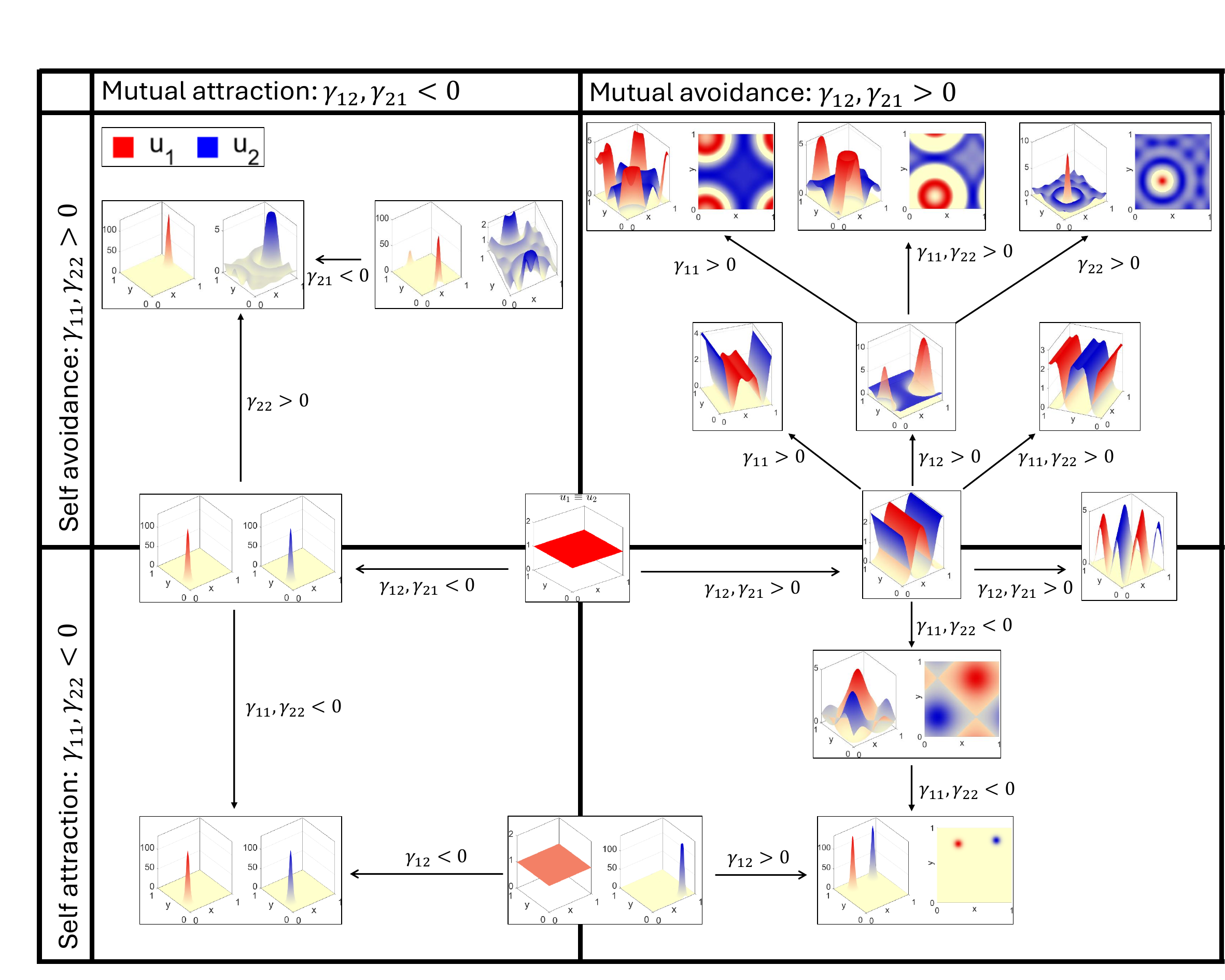}} 
         \caption{The phylogeny of pattern formation,  illustrating the role of interaction parameters $\gamma_{ij}$ in shaping the wide variety of emergent patterns. The figure presents stationary solutions, with arrows indicating transitions between different patterns as parameters are varied. Next to each arrow, we specify the parameter that is varied to move from one solution to another, with the arrow direction representing increasing magnitude of parameter values.}
         \label{fig:figure5}
\end{figure}

Having looked at a variety of subcases, we now summarise our various observations in the promised `phylogenic tree' of patterns, displayed in Figure \ref{fig:figure5}. This gives a single visual representation of how patterns emerge and morph as parameters are changed. For ease of digestion, this figure shows only stationary solutions, with arrows indicating transitions between different patterns as parameters are varied. Next to each arrow, we specify the parameter whose magnitude is increased to move from one solution to another, together with its sign. 

Our analysis highlights the crucial role of both self and mutual interaction parameters in shaping the spatial organisation of the two populations. The self-interaction parameters ($\gamma_{11}$, $\gamma_{22}$) primarily determine whether a population aggregates or disperses. When self-attraction is strong ($\gamma_{ii} < 0$), populations tend to cluster, regardless of whether mutual interactions favour attraction or avoidance. Conversely, self-repulsion ($\gamma_{ii} > 0$) favours dispersion, counteracting any tendency to cluster. In cases where self-repulsion is strong, patterns only emerge when mutual interactions are sufficiently large - either through strong mutual attraction pulling populations together, or strong mutual repulsion driving them apart. In addition to promoting dispersion, self-repulsion also introduces complex structures into the emerging solutions. For example, we observe the formation of ridges in striped patterns, waves in dispersed populations, and holes in aggregation solutions. These features appear in both mutual attraction and mutual avoidance regimes, as populations attempt to disperse and avoid excessive local density.

The mutual interaction parameters ($\gamma_{12}$, $\gamma_{21}$) further refine the organisation of the system. Mutual attraction ($\gamma_{12}, \gamma_{21} < 0$) consistently leads to aggregation, with both populations clustering in the same place. This solution is observed under a wide range of conditions and persists even in the presence of self-repulsion, provided the attraction is sufficiently strong. In contrast, mutual repulsion ($\gamma_{12}, \gamma_{21} > 0$) favours segregation, with populations forming distinct clusters or stripes. The specific structure depends on self-interaction: strong self-attraction leads to sharp segregation peaks, while weaker self-attraction results in rhombic clusters. When self-repulsion is included, the populations arrange themselves into striped formations. However, when the two mutual avoidance parameters are highly asymmetric ($\gamma_{12} \neq \gamma_{21}$), the system shows a clear shift in behaviour. Instead of forming symmetric segregation patterns, one population aggregates while the other disperses into the remaining space. Specifically, when $u_1$ strongly avoids $u_2$ ($\gamma_{12} \gg \gamma_{21}$), while $u_2$ is only weakly repelled or attracted by $u_1$, we observe that $u_1$ forms a compact, well-defined cluster, while $u_2$ disperses to occupy the rest of the domain. This behaviour highlights how avoidance-driven aggregation can arise in response to minimising interaction with another species, rather than self-attraction alone.

Beyond these static structures, the inclusion of a pursuit and avoidance scenario, where the mutual interaction parameters have opposite signs, gives rise to dynamic behaviours. Again, self-interaction plays a key role, as these dynamics emerge only when self-attraction is strong enough to sustain  aggregation. In regions where linear stability analysis predicts oscillations, transient dynamics can evolve into persistent oscillations, stationary segregation patterns or chase-and-run. Overall, our results show how a small set of interaction parameters can generate a rich diversity of spatial patterns and dynamic behaviours. The balance between self-attraction, self-repulsion and mutual interactions determines whether populations mix, segregate, oscillate or exhibit more complex formations such as ridges, holes or transverse stripes.

\section{Connecting theoretical patterns with biological observations}
The patterns observed in our study closely resemble many structures found in natural systems. These similarities suggest that  nonlocal interactions may play a fundamental role in shaping biological patterns.
In this section, we compare the emergent patterns of System \eqref{eq:model} with observed structures in nature. Rather than focusing solely on visual similarities, we make connections based on the underlying mechanisms at play. In particular, we draw parallels between the role of the nonlocal interaction parameters identified in our analysis and the known processes that govern self-organisation in biological systems.

\subsection{Mutual avoidance}
\subsubsection{Pedestrian lanes}
Pedestrian lane formation, in either humans or other mobile organisms, is an example of stripe formation driven by mutual avoidance. When two opposing streams of individuals cross, they instinctively avoid collisions by self-organising into alternating lanes \cite{helbing2005self}. This behaviour is often referred to as a smart collective pattern \cite{moussaid2012traffic}. Lane formation occurs when the density of the two streams is high enough for mutual repulsion to become sufficiently strong, leading to the spontaneous formation of parallel lanes \cite{feliciani2016empirical}. In these systems, the orientation of the lanes is influenced by environmental constraints and the crossing angle between the two flows, which determines the optimal orientation for efficient flow \cite{mullick2022analysis}.
This emergent pattern is very similar to the striped structures observed in our model, where mutual avoidance leads to spatial segregation (see, for example, Figure \ref{fig:Case1} or Figure \ref{fig:Case2} central grid). However, a key difference lies in the role of external directionality: in pedestrian systems, lane orientation is dictated by the imposed direction of movement, whereas in our model, stripe orientation emerges purely from nonlocal interactions, without any predefined directional cues. Our results suggest that mutual avoidance alone is a sufficient mechanism to drive lane formation, even in the absence of external cues. Their orientation might then be shaped by additional factors such as direction of movement or environmental constraints.

\subsubsection{Space use patterns in territorial animals.}
Territorial animals actively defend a specific territory from members of the same or different species. These territories are used for feeding, mating, or raising offspring, and their spatial organisation often reflects behavioural interactions and environmental pressures. In territorial animals, space use often results in contiguous or separated territorial patterns that can change dynamically in response to changes in population size or neighbourhood mortality. In  \cite{potts2013quantifying} the authors observed that in urban red fox populations, individuals maintain contiguous territorial boundaries even when neighbouring individuals disappear. This occurs because surviving individuals expand their territory into vacated areas, while maintaining overall territorial contiguity.  Similar behaviour emerges in our simulations: when mutual avoidance is present, we observe the formation of contiguous spatial patterns, as also previously in related models \cite{potts2016memory,potts2016territorial}. These solutions can take the form of \textit{Diamond Patterns} when there is weak self-attraction (see Figure \ref{fig:Case4}, central grid), or \textit{Stripes} when there is no self-interaction or weak self-avoidance (see Figures \ref{fig:Case2} and \ref{fig:Case3}, central grids). In these cases, populations organise themselves into well-defined regions of space; they are not segregated, but show overlap between adjacent territories. 

When mutual repulsion becomes strong, \textit{Diagonal Stripes} emerge, consisting of segregated clusters with no overlap between populations (see Figure \ref{fig:Case2} , central grid, $(\gamma_{21},\gamma_{12})=(5,5)$). Similar spatial arrangements are found in nature. For example, studies of wolf populations have shown the presence of buffer zones between adjacent packs, where direct encounters are rare and prey species often reside \cite{lewis1993modelling}. As noted in \cite{lewis1993modelling}, such territories are typically arranged in a spatial mosaic, with clearly defined boundaries between neighbouring groups.

\begin{figure}
       \centering  {\includegraphics[width=1\textwidth]{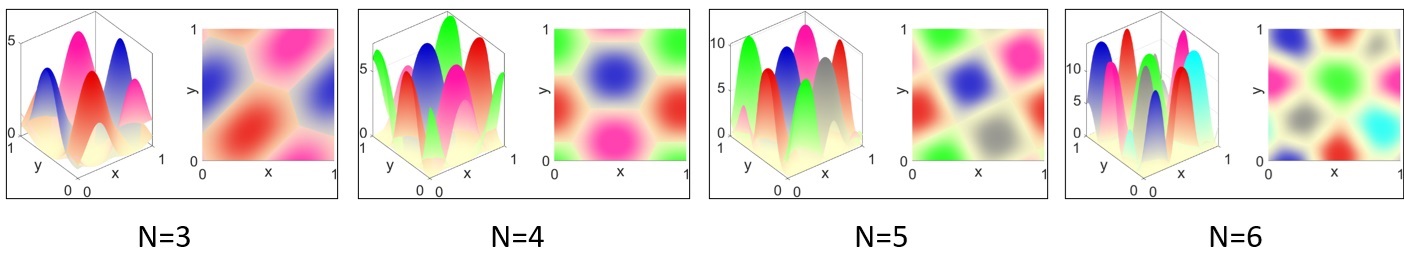}} 
       \caption{Geometric stationary patterns obtained by solving System \eqref{eq:model}, with $K_{ij}$ given in \eqref{eq:kernel}, for $N=3,4,5,6$ species. The other parameter values are: $D_i=1$, $R_{ij}=R_{ji}=0.2$, $m_i=1,$ $\gamma_{ij}=\gamma_{ji}=1.2$, if $i\neq j$, and $\gamma_{ii}=0$, for $i,j=1,\dots,N$. The initial condition is a small random perturbation of the homogeneous steady state. The figure shows the state of the system after the transient dynamics have subsided.}
       \label{fig:MUltispecies}
\end{figure}

To explore the phenomenon of territory formation further, we examine emerging patterns as the number of populations increases.
To this end, we performed a numerical investigation for $\gamma_{ij}=\gamma_{ji} >0$ (mutual avoidance) for  $N=3,4,5,6$ populations, and the results are shown in Figure \ref{fig:MUltispecies}. With three and four populations, we observe hexagonal tilings. Also, we find regular tessellations with squares for $N=5$, and irregular but still spatially well-defined domains for $N=6$. These stationary patterns show marked similarities to territorial structures observed in nature. The formation of territorial patterns with well-defined edges, often resembling polygons, has been the subject of both theoretical and empirical studies \cite{adams2001approaches}.  Theoretical arguments from geometry and dynamical models predict that high intraspecific antagonism in dense populations living in homogeneous environments should promote hexagonal distributions \cite{pringle2017spatial}. Empirical evidence supporting this hypothesis has been documented in \cite{barlow1974hexagonal}, where it was observed polygonal territories in the cichlid fish \textit{Tilapia mossambica}, characterised by five or six clearly defined ridges. In addition, analysis of territorial data on the pectoral sandpiper (\textit{Calidris melanotos}), has revealed that adjacent sides of territories can meet at an average interior angle of 117° \cite{grant1968polyhedral}, which is close to the 120° angle characteristic of hexagonal tilings.

Our model provides a theoretical framework that predicts how nonlocal interactions can drive the emergence of polygonal territorial structures, mirroring those observed in natural systems.  These results suggest that nonlocal mechanisms may play a fundamental role in shaping the geometry of territories. As such, this class of models provides a useful tool for investigating how social behaviour and spatial dynamics interact to influence the formation and maintenance of territories. Understanding these processes is indeed essential for interpreting spatial population patterns, managing wildlife habitats, and informing conservation strategies where territoriality plays a key ecological role.

\subsection{Self-avoidance}
\subsubsection{Spatial distribution of resident macrophages}
Our numerical simulations show that sufficiently strong self-avoidance produces distinct modulations in the spatial patterns. For example, when self-avoidance is strong, stripe patterns develop pronounced ridges (see Figure \ref{fig:Case1}, $(\gamma_{ii},\gamma_{ij})=(3,5)$), nearly homogeneous distributions exhibit waves or spot formations (see Figure \ref{fig:Case2}, left grid, $(\gamma_{21},\gamma_{12})=(4,4)$), and clusters can form central voids resembling volcanic structures (see Figure \ref{fig:Case3}, left grid, $(\gamma_{21},\gamma_{12})=(4,5)$). 
We also find similar patterns in biological systems, perhaps driven by similar mechanisms. Specifically, distributions of resident macrophages (a family of immune cells that act as sentinels in almost every organ), exhibit a regular spacing and corridors (see \cite{hume2019mononuclear} Figure 1) that mirrors \textit{Ridged Stripes} or \textit{Co-Aggregates with halo}. Research suggests that such regular spacing is due to self-avoidance \cite{avetisyan2018muscularis}: macrophages actively survey large areas of their environment using highly motile filopodia and, when encountering neighbouring cells, initiate repulsive responses.  This effectively establishes distinct cell territories in a cell-autonomous manner. Although additional factors such as tissue-specific factors or inflammatory status can also influence cellular organisation, mutual repulsion appears to be a key driver \cite{sreejit2020origins}. Thus, our model is able to reproduce patterns generated by self-repulsion, thus capturing fundamental principles underlying spatial organisation in biological systems.
 
Resident macrophages, in addition to their role in removing harmful substances, also contribute to the amplification of tissue inflammation \cite{ jaeschke1997mechanisms}. Concentric ring patterns are a hallmark of many inflammatory diseases, including the brain plaques of multiple sclerosis \cite{lassmann2005multiple} and the ring- and bull's-eye-shaped rashes characteristic of erythema anulare centrifugum \cite{ ingen2017diagnosis}. Reaction-diffusion-chemotaxis models have successfully reproduced such patterns \cite{ lombardo2017demyelination, bilotta2019axisymmetric, giunta2021pattern}, linking them to the complex interplay between cell movement and chemical signalling. Our model provides a complementary perspective by suggesting that mutual avoidance between macrophages may drive the formation of these rings. 
\subsubsection{Bacteria distribution}

In \cite{ woodward1995spatio}, a study of Salmonella Typhimurium reports the formation of continuous and perforated concentric rings (see \cite{ woodward1995spatio} Figure 1), which closely resemble the perforated rings observed in \textit{Splash Patterns}. In \cite{ woodward1995spatio}, a chemotaxis model was used to reproduce the observed ring formations. Our analysis provides an alternative hypothesis: that holes and perforated rings may emerge as an intrinsic mechanism to alleviate high local densities within the bacterial population. This suggests that even in the absence of chemotactic cues, the tendency of bacteria to repel each other can drive the formation of comparable spatial structures. Our numerical findings can not only complement existing chemotaxis-based explanations, but also provide a valuable tool for further investigating the role of self-avoidance in biological pattern formation.

\subsection{Self-attraction}

\subsubsection{Cell Sorting}

In a series of classic studies, Steinberg \cite{Steinberg1963,Steinberg1970} investigated how cells of different adhesive strength rearrange themselves in co-cultures. In their experiments, two types of embryonic chick cells - identical except for their adhesive properties - were mixed in a Petri dish. Type A cells adhered strongly to each other, while type B cells adhered more weakly. Depending on the relative strength of self- and cross-adhesion between the two types, a variety of spatial arrangements emerged.

Many of the patterns observed in these cell sorting experiments are very similar to those produced by our model. In particular, configurations resembling \textit{Segregated Clusters}, \textit{Diamonds}, \textit{Stripes}, and \textit{Co-Aggregates} were reported in \cite{Steinberg1970} and subsequently reproduced using nonlocal partial differential equation models by \cite{armstrong2006continuum} and \cite{carrillo2019population}. These studies quantified the adhesive interactions required to reproduce each configuration and provided a theoretical framework for interpreting the experimental results. Similarly, in our model, variations in the strengths of self- and mutual interactions lead to qualitatively analogous spatial arrangements.

\subsubsection{Metastasis} 

Metastatic cells are known to lose their adhesion from the primary tumor, allowing them to detach from the primary tumour, disperse to other parts of the body, and form new aggregates in distant tissues \cite{martin2013cancer}. 
In our simulations, we observe that reducing the self-attraction within one population leads to its dispersal into the surrounding space. For example, comparing the case with $\gamma_{11} = -1$ and $\gamma_{22} = -1$ (see Figure \ref{fig:Case4}, middle grid) with the case with $\gamma_{11} = 0$ and $\gamma_{22} = -1$ (see Figure \ref{fig:Case2}, right grid), we see that under mutual avoidance conditions, removing self-attraction in $u_1$ causes it to disperse into the space not occupied by $u_2$. Similarly,
if we ignore mutual interactions and consider the case where $\gamma_{12}=\gamma_{21}=0$ and $\gamma_{11}=\gamma_{22}=-2$ (see Figure \ref{fig:Case1}), we observe the formation of \textit{Segregated Clusters}. Reducing the self-adhesion in $u_1$ (i.e. by setting $\gamma_{11} = -1$) while keeping all other parameters fixed (see Figure \ref{fig:Case4}, right grid) leads to the dispersion of $u_1$, while $u_2$ remains aggregated. These observations suggest that the loss of self-attraction is the primary driver of dispersal, reflecting the dynamics that underlie the initial stages of metastasis.

\section{Discussion}
We have performed an extensive numerical investigation of a 2-species nonlocal advection-diffusion model on a 2D domain, revealing a diverse range of self-organised spatial patterns arising from different combinations of self- and mutual interactions. By systematically varying key parameters governing nonlocal attraction and repulsion, we have identified how each interaction mechanism contributes to the emergence of specific structures, leading to a `phylogenetic tree' of pattern formation in such systems (Figure \ref{fig:figure5}). Many of the spatial structures observed in our simulations resemble patterns found in natural systems, suggesting a wide range of possible applications in biology, ecology, and social sciences. 

Our results show that complex spatial patterns can emerge purely from movement-driven interactions. This underscores the potential of nonlocal models as minimal yet powerful frameworks for understanding pattern formation. The combination of linear stability analysis and simulations provides insight into both the emergence of instabilities and the selection of stationary or dynamic patterns. The results of this study serve as a foundation for further theoretical and numerical research, guiding future investigations into more complex scenarios.

From a mathematical point of view, the variety of patterns and transitions observed numerically indicates a rich bifurcation structure underlying the dynamics of the system. Understanding these bifurcations - including their onset, type and stability properties - is crucial for predicting which patterns may emerge under given biological conditions. Analytical techniques such as  spectral and weakly nonlinear analysis \cite{giunta2024weakly} can provide local insights near bifurcation points, while numerical continuation methods (e.g. those implemented in software packages such as pde2path \cite{uecker2014pde2path} or AUTO \cite{doedel2010auto}) can track solution branches in the nonlinear regime. These tools can help to construct detailed bifurcation diagrams that not only classify the possible solution types, but also help to identify parameter regimes where patterns are robust or particularly sensitive to perturbations \cite{soresina2022hopf}. 
Such understanding has direct implications for applications such as cancer modelling, where distinguishing the stability of clustered versus dispersed configurations can inform hypotheses about tumour invasion mechanisms.

A promising extension of our study is a deeper exploration of multi-species dynamics beyond the two-species framework. Biological systems often consist of multiple interacting populations with different interaction strengths, diffusion rates and sensing ranges. Generalising the model to include more species could reveal new types of emergent patterns. Such studies could be particularly relevant for understanding biodiversity in ecosystems, multi-cellular organisation in tissues, or the spatial dynamics of social groups in human systems. From a modelling perspective, our framework can naturally be extended to include additional biologically-motivated features. These might include anisotropic sensing or movement (e.g. in response to gradients), domain heterogeneity, and interactions with boundaries or obstacles.

In summary, our work highlights the fundamental role of nonlocal interactions in driving pattern formation and shows how simple interaction rules can lead to self-organisation. By connecting mathematical analysis with biological examples, this study opens new avenues for the application of nonlocal models to a wide range of systems, offering both theoretical insights and possible directions for future empirical investigations.\\

\bigskip

\enlargethispage{20pt}








\printbibliography
\end{document}